\documentclass[a4wide,10pt]{article}

\newcommand{\vb}[1]{{\mathbf{#1}}}
\newcommand{\lb}[1]{\label{#1}}

\newcommand{\bc}{\begin{center}}
\newcommand{\ec}{\end{center}}
\newcommand{\be}{\begin{equation}}
\newcommand{\ee}{\end{equation}}
\newcommand{\bea}{\begin{eqnarray}}
\newcommand{\eea}{\end{eqnarray}}
\newcommand{\ba}[1]{\begin{array}{#1}}
\newcommand{\ea}{\end{array}}

\newcommand{\bt}[1]{\begin{table}[ht]\centering\begin{tabular}{#1}}
\newcommand{\et}[1]{\end{tabular}\caption{\small#1}\end{table}}

\begin{document}

\begin{titlepage}
\title{Effective Electric and Magnetic Local Actions for\\[3mm] $U_e(1)\times U_g(1)$ Electromagnetism:\\[3mm] Hodge Duality and Zero-Field Equation}

\author{Pedro Castelo Ferreira\\[3mm] \small CENTRA, Instituto Superior T\'ecnico, Av. Rovisco Pais, 1049-001 Lisboa, Portugal}

\label{firstpage}

\maketitle

\begin{abstract}{Electromagnetism, Dynamical Symmetry Breaking, pseudo-photons, Zero Field Equation; packs: 03.50.De,11.15,-q,40.}
In this work it is considered a mechanism of dynamical symmetry breaking for extended $U_e(1)\times U_g(1)$
containing, one vector gauge field $A$ (photon) and one pseudo-vector gauge field $C$ (pseudo-photon).
By choosing a particular solution of the equations of motion we obtain a functional description of either field
in terms of the other one. In this way we obtain non-trivial configurations $C=C(A)$ or $A=A(C)$ such that, in
the effective broken theories containing only one gauge field, the usual field discontinuities in the presence of both electric
and magnetic monopoles (Dirac string or Wu-Yang non-trivial fiber-bundle) are absent. These fields regularity is achieved
through corrections, inherited from the unbroken theory, to the standard definitions of electromagnetic fields and four-currents.
Based in these results we also demonstrate non-triviality
of the unbroken theory, in the presence of both electric and magnetic four-currents, there are no trivial solutions compatible
with the equations of motion for both fields. Moreover we demonstrate that, although in complete agreement with
the Maxwell equations, extended $U_e(1)\times U_g(1)$ electromagnetism is not equivalent to the standard variational $U(1)$
Maxwell electromagnetism. We further show that the unbroken theory is invariant under a combination of Hodge
dualities for the gauge connections, $F\to\hat{\epsilon}*G$ and $G\to-\hat{\epsilon}*F$ ($\hat{\epsilon}=\pm 1$), that has
as self-dual point, the field configuration corresponding to the zero-field equation $\tilde{C}=-\hat{\epsilon}A$. In addition this
condition has the particularity of being the only configuration compatible with the mechanism of dynamical symmetry breaking proposed
here that is also gauge invariant in the unbroken theory. These characteristics justify the zero-field equation as being a
preferred configuration.
\end{abstract}

\end{titlepage}

\section{Introduction}

The inclusion of magnetic charge in the Maxwell equations is justified by being the only theoretical explanation for
quantization of electric charge due to the Dirac quantization condition $eg=n$~(Dirac). However the introduction
of magnetic charge in single $U(1)$ photon theories implies the existence of non-physical extended singularities known as the Dirac
string~(Dirac) or the Wu-Yang fiber bundle~(Wu \& Yang~1975). These singularities are due to the violation of the
Bianchi identities for the gauge field along a line (string) or a plane (the gluing of the bundles).
A possible approach that eliminates extended singularities is to consider one extra auxiliary gauge field $C$.
This framework has been originally put forward by Cabibbo and Ferrari~(Cabibbo \& Ferrari 1962) and further developed by Schwinger~(Schwinger~1966).
More recently, the extension to a theory with gauge group $U_e(1)\times U_g(1)$, containing both a photon $A$ and a pseudo-photon $C$,
have been studied in~(Singleton~1995,Cardoso de Mello \& al.~1996, Berkovits~1996, Carneiro~1997, Castelo Ferreira~2005)
and hold regular field solutions in the presence of both
electric and magnetic monopoles. Independently of the existence of magnetic monopoles the inclusion of a
pseudo-photon is also justified in the presence of external non-trivial field configurations that violate
the gauge fields Bianchi identities which, at variational level, cannot be accounted by the Maxwell action
for a single photon~(Castelo Ferreira~2006). Theories with a pseudo-photon $C$ raise also the question to whether this
field is simply a mathematical artifact, or a real physical particle that have so far not been directly detected.
If the pseudo-photon is a real physical field it remains the pertinent question weather such theory has some sort of phase transition and
which physical systems are in an unbroken phase, where both $A$ and $C$ fields are present,
or in a broken phase of the theory were only the standard photon or pseudo-photon are present.
Already available theoretical results, indicate that exist both physical systems which exhibit an unbroken
phase (electromagnetism in the presence of non-regular external electromagnetic fields~(Castelo Ferreira~2006)), physical
systems which exhibit a broken electric phase (Anderson-Schwinger mechanism~(Schwinger~1962, Anderson~1963, Proca~1988) for plasmon mass generation~(Castelo Ferreira \& Mendon\c{c}a~2006, Mendon\c{c}a \& Castelo Ferreira~2006)) and systems that exhibit a broken magnetic phase (fractional Hall effect~(Tsui \& al.~1982, Laughlin~1982, Girvin \& MacDonald~1987, Zhang \& al.~1989, Jain~1989) in planar systems~(Castelo Ferreira~2007)).
These studies indicate that the pseudo-photon may be a truly physical field, instead of a mathematical auxiliary field.
Also it is important to note that in~(Castelo Ferreira~2007) is proved the equivalence between Dirac's quantization condition and
the experimental measured quantization of magnetic flux in the fractional Hall effect. This equivalence is only fully consistent
in extended $U_e(1)\times U_g(1)$ electromagnetism. These results motivate further studies in this theory.

The action with gauge symmetry $U_e(1)\times U_g(1)$ for electromagnetism, with one gauge vector field $A$
and one gauge pseudo-vector field $C$, coupled to electric and magnetic sources is given
by $S=S_0+S_{\mathrm{Sources}}$~(Cardoso de Mello \& al.~1996, Castelo Ferreira~2005),
\be
S_0=-\int_M\left[\frac{1}{4}F_{\mu\nu}F^{\mu\nu}-\frac{1}{4}G_{\mu\nu}G^{\mu\nu}-\frac{\hat{\epsilon}}{4}\epsilon^{\mu\nu\lambda\rho}F_{\mu\nu}G_{\lambda\rho}\right]\ ,
\lb{S_Unbroken_0}
\ee
where the gauge connections are $F_{\mu\nu}=\partial_\mu A_\nu-\partial_\nu A_\mu$ and $G_{\mu\nu}=\partial_\mu C_\nu-\partial_\nu C_\mu$,
and 
\be
S_{\mathrm{Sources}}=-\int_M\left[(A_\mu-\hat{\epsilon}\tilde{C}_\mu) J_e^\mu-(\hat{\epsilon}C_\mu+\tilde{A}_\mu) J_g^\mu\right]\ .
\lb{S_Unbroken_J}
\ee
$\tilde{A}$ and $\tilde{C}$ do not constitute independent fields and are defined by the
differential equations~(Castelo Ferreira~2006)
\be
\tilde{F}^{\mu\nu}=\frac{1}{2}\epsilon^{\mu\nu\lambda\rho}F_{\lambda\rho}\ \ , \ \ \tilde{G}^{\mu\nu}=\frac{1}{2}\epsilon^{\mu\nu\lambda\rho}G_{\lambda\rho}\ ,
\lb{dual_fields}
\ee
where the dual gauge connections are $\tilde{F}_{\mu\nu}=\partial_\mu \tilde{A}_\nu-\partial_\nu \tilde{A}_\mu$ and 
$\tilde{G}_{\mu\nu}=\partial_\mu \tilde{C}_\nu-\partial_\nu \tilde{C}_\mu$.
The electromagnetic physical fields in the unbroken theory corresponding to the above action are defined as~(Singleton~1995, Castelo Ferreira~2005)
\be
E^i= F^{0i}-\frac{\hat{\epsilon}}{2}\epsilon^{0ijk}G_{jk}\ \ ,\ \ B^i=\hat{\epsilon}G^{0i}+\frac{1}{2}\epsilon^{0ijk}F_{jk}\ .
\lb{ph_fields_FG}
\ee
In the original works~(Cabibbo \& Ferrrari~1962, Schwinger~1966) the pseudo-photon was considered to be a non-physical auxiliary field,
which through an appropriate constraint, would be effectively excluded from the theory. The specific constraint considered
in these original works, as well as in subsequent works~(Cardoso de Mello \& al.~1996, Berkovits~1996, Carneiro~1997) is the zero-field condition
\be
G^{\mu\nu}=\frac{\hat{\epsilon}}{2}\epsilon^{\mu\nu\lambda\rho}F_{\lambda\rho}\ \ \Leftrightarrow\ \ \tilde{C}_\mu=-\hat{\epsilon} A_\mu\ .
\lb{0-field}
\ee
Alternative approaches were further developed in~(Zwanzinger~1968, Brandt \& al.~1979), with the drawback of not preserving either space isotropy or Lorentz invariance,
as well as to consider a very massive pseudo-photon~(Singleton~1995). In this work our aim is to investigate if
effective theories with one single gauge field can be obtained from a theory with photons and pseudo-photons
through some mechanism of dynamical symmetry breaking. In particular if the zero-field equation~(\ref{0-field})
is applicable in this framework and what are the consequences of such mechanism at the level of the effective
broken theories. We are going to work mainly at variational level, having as starting point the above action
$S=S_0+S_{\mathrm{Sources}}$ given in~(\ref{S_Unbroken_0}) and~(\ref{S_Unbroken_J}), and using the equations
of motion to find appropriate field configurations.

\section{\lb{sols}Effective Theories in the Absence of Sources}

In this section we derive the equations of motion for action~(\ref{S_Unbroken_0}) and
show that the generic non-trivial solutions kill half of the degrees of freedom of the full theory
rendering either the standard Maxwell action, with only one gauge field, or its magnetic counterpart.
As we will show both effective actions differ by a minus sign such that the pseudo-vector gauge field is a ghost
field, as already shown in~(Castelo Ferreira~2006).

For non-regular gauge fields the equations of motion obtained by varying the action~(\ref{S_Unbroken_0}) with respect to
$C_\nu$ and $A_\nu$ are
\be
\partial_\mu G^{\mu\nu}=-\frac{\hat{\epsilon}}{2}\epsilon^{\mu\nu\lambda\rho}\partial_\mu F_{\lambda\rho}\ \ ,\ \ 
\partial_\mu F^{\mu\nu}=\frac{\hat{\epsilon}}{2}\epsilon^{\mu\nu\lambda\rho}\partial_\mu G_{\lambda\rho}\ .
\lb{EOMFG_nonreg}
\ee
We note that these two equations are linear dependent, in particular are Hodge conjugated to each other, hence they constitute
only 4 independent equations corresponding to $\nu=0,1,2,3$. Also a direct conclusion from the above equations is that
the Bianchi identities for each field are related to the equations of motion for the other field. It is this fact that in the presence
of external non-regular electromagnetic fields allow the induced electromagnetic fields to be expressed in terms of regular
gauge fields only. External fields have been studied in detail in~(Castelo Ferreira~2006), in the present study we consider only external
electric and magnetic four-currents. We will first address the case of non-regular fields and later address the case of regular fields.

\subsection{Electric Solutions for Non-Regular Fields}

The non-trivial generic solutions for the equations of motion~(\ref{EOMFG_nonreg}) are obtained by direct integration of the equations
and are defined up to a closed 2-form, i.e. an antisymmetric tensor $f_{\mu\nu}$, defined in terms of a regular field $a_\mu$,
\be
G^{\mu\nu}=-\frac{\hat{\epsilon}}{2}\epsilon^{\mu\nu\lambda\rho}F_{\lambda\rho}+\frac{\alpha_e}{2}\epsilon^{\mu\nu\lambda\rho}f_{\lambda\rho}\ \ ,\ \ 
f_{\mu\nu}=\partial_\mu a_\nu-\partial_\nu a_\mu\ \ ,\ \ \epsilon^{\mu\nu\lambda\rho}\partial_\nu f_{\lambda\rho}=0\ .
\lb{G_nonreg_0}
\ee
In these solutions we have considered, for convenience, a dimensionless constant that, up to rescaling of the fields $a_\mu$, can be set to unity
$\alpha_e=\pm 1$. By replacing the solution~(\ref{G_nonreg_0}) in the original action~(\ref{S_Unbroken_0}), we obtain for each of
the terms constituting the action, the following expressions
\be
\ba{rcl}
\displaystyle+\frac{1}{4}F_{\mu\nu}F^{\mu\nu}&=&\displaystyle+\frac{1}{4}F_{\mu\nu}F^{\mu\nu}\ ,\\[2mm]
\displaystyle-\frac{1}{4}G_{\mu\nu}G^{\mu\nu}&=&\displaystyle+\frac{1}{4}F_{\mu\nu}F^{\mu\nu}+\frac{\alpha_e^2}{4}f_{\mu\nu}f^{\mu\nu}-\frac{2}{4}\hat{\epsilon}\alpha_eF_{\mu\nu}f^{\mu\nu}\ ,\\[2mm]
\displaystyle-\frac{\hat{\epsilon}}{4}\epsilon^{\mu\nu\lambda\rho}F_{\mu\nu}G_{\lambda\rho}&=&\displaystyle-\frac{2}{4}F_{\mu\nu}F^{\mu\nu}+\frac{2}{4}\hat{\epsilon}\alpha_eF_{\mu\nu}f^{\mu\nu}\ .
\ea
\lb{terms_e_nonreg}
\ee
Here we used the identity 
\be
\epsilon^{\mu\nu\delta\rho}X_{\delta\rho}\epsilon_{\mu\nu}^{\ \ \ \delta'\rho'}Y_{\delta'\rho'}=-4X_{\mu\nu}Y^{\mu\nu},
\lb{HH_id}
\ee
valid for antisymmetric rank two tensors $X$ and $Y$ in a flat $3+1$--dimensional Minkowski space.
Replacing the expressions~(\ref{terms_e_nonreg}) in the unbroken action~(\ref{S_Unbroken_0}) and electromagnetic
field definitions~(\ref{ph_fields_FG}) we obtain the effective action and electromagnetic field
definitions
\be
S_{\mathrm{Electric}}=-\int_M \frac{1}{4}f_{\mu\nu}f^{\mu\nu}\ \ ,\ \ E^i=\hat{\epsilon}\alpha_ef^{0i}\ \ ,\ \ B^i=\frac{\hat{\epsilon}\alpha_e}{2}\epsilon^{0ijk}f_{jk}\ .
\lb{el_eff_0}
\ee
This action is recognized as the standard Maxwell action for the gauge field $a_\mu$ with gauge symmetry $U_{\bar{e}}(1)$ and,
for $\hat{\epsilon}\alpha_e=+1$, the field definitions correspond to the standard ones of electromagnetism. The equations of motion
for the broken theory are now given by
\be
\partial_\mu f^{\mu\nu}=0\ .
\lb{EOM_f}
\ee
These equations can consistently be obtained by direct replacement of the solution~(\ref{G_nonreg_0}) in the equation of motion~(\ref{EOMFG_nonreg}),
or by a variation of the effective broken action~(\ref{el_eff_0}) with respect to the field $a_\nu$.
The equations of motion for $a$~(\ref{EOM_f}), together with the respective Bianchi identities~(\ref{G_nonreg_0}),
constitute the standard Maxwell equations.

\subsection{Magnetic Solutions for Non-Regular Fields}

To obtain the effective magnetic solution we use the same approach. The non-trivial solution of~(\ref{EOMFG_nonreg}) for $F_{\mu\nu}$
is, up to a generic antisymmetric tensor $g_{\mu\nu}$ defined in terms of a regular gauge field $c_\mu$, given by
\be
F^{\mu\nu}=\frac{\hat{\epsilon}}{2}\epsilon^{\mu\nu\lambda\rho}G_{\lambda\rho}+\frac{\alpha_g}{2}\epsilon^{\mu\nu\lambda\rho}g_{\lambda\rho}\ \ ,\ \ 
g_{\mu\nu}=\partial_\mu c_\nu-\partial_\nu c_\mu\ \ ,\ \ \epsilon^{\mu\nu\lambda\rho}\partial_\nu g_{\lambda\rho}=0\ .
\lb{F_nonreg_0}
\ee
Again, for convenience, we considered a constant $\alpha_g=\pm 1$. Replacing this solution in the original
unbroken action~(\ref{S_Unbroken_0}), we obtain the following expressions for each of the terms
\be
\ba{rcl}
\displaystyle-\frac{1}{4}G_{\mu\nu}G^{\mu\nu}&=&\displaystyle-\frac{1}{4}G_{\mu\nu}G^{\mu\nu}\ ,\\[2mm]
\displaystyle+\frac{1}{4}F_{\mu\nu}F^{\mu\nu}&=&\displaystyle-\frac{1}{4}G_{\mu\nu}G^{\mu\nu}-\frac{\alpha_g^2}{4}g_{\mu\nu}g^{\mu\nu}-\frac{2}{4}\hat{\epsilon}\alpha_gG_{\mu\nu}g^{\mu\nu}\ ,\\[2mm]
\displaystyle-\frac{\hat{\epsilon}}{4}\epsilon^{\mu\nu\lambda\rho}F_{\mu\nu}G_{\lambda\rho}&=&\displaystyle+\frac{2}{4}G_{\mu\nu}G^{\mu\nu}+\frac{2\hat{\epsilon}\alpha_g}{4}G_{\mu\nu}g^{\mu\nu}\ ,
\ea
\lb{terms_g_nonreg}
\ee
where again we used the identity~(\ref{HH_id}). Hence from~(\ref{S_Unbroken_0}) we obtain the effective action, and from~(\ref{ph_fields_FG}) the electromagnetic fields definitions
\be
S_{\mathrm{Magnetic}}=+\int_M \frac{1}{4}g_{\mu\nu}g^{\mu\nu}\ \ ,\ \ E^i=\frac{\alpha_g}{2}\epsilon^{0ijk}g_{jk}\ \ ,\ \ B^i=-\alpha_gg^{0i}\ .
\lb{mg_eff_0}
\ee
This action has the opposite sign than the usual Maxwell action. At classical level this is not relevant, however upon quantization
it renders negative energy eigenstates, thus the field $c_\mu$ is interpreted as a ghost field~(Castelo Ferreira~2006) with gauge symmetry $U_{\bar{g}}(1)$.
By comparing these definitions with the original field definitions~(\ref{ph_fields_FG}) for the unbroken theory, we conclude that for
$\alpha_g=-\hat{\epsilon}$ are obtained the standard field definitions. The equations of motion are given by
\be
\partial_\mu g^{\mu\nu}=0\ ,
\lb{EOM_g}
\ee
and are obtained consistently, either by direct replacement of the solution~(\ref{F_nonreg_0}) in the equation of motion~(\ref{EOMFG_nonreg}),
or by a variation of the effective broken action~(\ref{mg_eff_0}) with respect to the field $c_\nu$.
The equations of motions for $c$~(\ref{EOM_g}), together with the respective Bianchi identities~(\ref{F_nonreg_0}),
constitute the magnetic counterpart of the Maxwell equations.

\subsection{Solutions for Regular Fields}

For regular gauge fields the Hopf term in the action~(\ref{S_Unbroken_0}) is a total derivative and does not contribute
to the equations of motion. Therefore, upon variation of the action~(\ref{S_Unbroken_0}) with respect 
to the gauge fields $C_\nu$ and $A_\nu$, we obtain the equations
\be
\partial_\mu G^{\mu\nu}=0\ \ ,\ \ \partial_\mu F^{\mu\nu}=0\ .
\lb{EOMFG_reg_0}
\ee
Due to the fields being regular these equations are supplemented by the respective Bianchi identities
\be
\epsilon^{\mu\nu\lambda\rho}\partial_\mu G_{\lambda\rho}=0\ \ ,\ \ \epsilon^{\mu\nu\lambda\rho}\partial_\mu F_{\lambda\rho}=0\ .
\lb{BI_FG}
\ee
The most generic solution for both equations of motion~(\ref{EOMFG_reg_0}) are, respectively,
\be
G^{\mu\nu}=\epsilon^{\mu\nu\delta\rho}h^e_{\delta\rho}\ \ ,\ \ F^{\mu\nu}=\epsilon^{\mu\nu\delta\rho}h^g_{\delta\rho}\ ,
\lb{FG_reg_h}
\ee
for generic rank two tensors obeying $\epsilon^{\mu\nu\delta\rho}\partial_\nu h^g_{\delta\rho}=\epsilon^{\mu\nu\delta\rho}\partial_\nu h^e_{\delta\rho}=0$. We are considering regular gauge fields only, hence without loss of generality, we can consider the field redefinitions
\be
h^g_{\mu\nu}=+\frac{\hat{\epsilon}}{2}G_{\mu\nu}+\frac{\alpha_g}{2}g_{\mu\nu}\ \ ,\ \ 
h^e_{\mu\nu}=-\frac{\hat{\epsilon}}{2}F_{\mu\nu}+\frac{\alpha_e}{2}f_{\mu\nu}\ ,
\lb{FG_reg_0}
\ee
with $f_{\mu\nu}$ and $g_{\mu\nu}$ defined as in~(\ref{G_nonreg_0}) and~(\ref{F_nonreg_0}). In this way
we retrieve the same expressions for the solutions that we have obtained for non-regular gauge fields in~(\ref{G_nonreg_0}) and~(\ref{F_nonreg_0}),
respectively. The remaining of the proof follows in the same manner as for non-regular gauge fields, such that we obtain the respective effective electric action~(\ref{el_eff_0}) and effective magnetic action~(\ref{mg_eff_0}). However it is important to stress that here, the solutions~(\ref{FG_reg_0}) for each gauge fields $C$ and $A$, are regular and the coupling of both sectors can only be fully justified by demanding
consistence of the theory with the existence of non-regular gauge fields~(Castelo Ferreira~2006) (or/and magnetic monopoles~(Castelo Ferreira~2006)).
To finalize our discussion we recall that the standard definitions of the electromagnetic fields and actions in electromagnetism
correspond to setting the integration constants $\alpha_e$ and $\alpha_g$ to be
\be
\alpha_e=\hat{\epsilon}\ \ ,\ \ \alpha_g=-\hat{\epsilon}\ .
\lb{signs_0}
\ee
Here $\hat{\epsilon}$ stands for the relative terms of the Hopf term in~(\ref{S_Unbroken_0}).

\subsection{On Trivial Solutions}

So far we have not addressed the trivial solutions of the equations of motion. In addition to the non-trivial solutions~(\ref{G_nonreg_0}) and~(\ref{F_nonreg_0}) we can consider the cases $\alpha_e=0$ and $\alpha_g=0$ which correspond to
\be
\tilde{C}_\mu=+\hat{\epsilon}A_\mu\ .
\lb{sols_trivial}
\ee
This solution holds a trivially null effective action $S_{\mathrm{eff}}=0$, which implies killing all the dynamics. Moreover, without any
external sources or external fields, this solution is actually expected. The theory does not have any dynamics and the
action is null. Hence the non-trivial solutions~(\ref{G_nonreg_0}) and~(\ref{F_nonreg_0}), although formally
obeying the equations of motion (being an extrema of the action), cannot be physically justified as preferred in relation
to the trivial solutions~(\ref{sols_trivial}). In the next section we include external source terms which, as we will show, justify the
non-trivial solutions as the only allowed ones, the trivial solution will no-longer be extrema of the action.

\section{\lb{sources}Inclusion of Source Terms}

We proceed now to compute the solutions of the equations of motion in the presence of both electric and magnetic 
four-current densities. For non-regular gauge fields, upon variation of the full action $S=S_0+S_{\mathrm{Sources}}$
given by~(\ref{S_Unbroken_0}) and~(\ref{S_Unbroken_J}) with respect to $C_\nu$ and $A_\nu$, we obtain the equations
\bea
\partial_\mu G^{\mu\nu}=\hat{\epsilon}J_g^\nu-\frac{\hat{\epsilon}}{2}\epsilon^{\mu\nu\lambda\rho}\partial_\mu F_{\lambda\rho}\ ,\lb{EOMG_nonreg_J}\\
\partial_\mu F^{\mu\nu}=J_e^\nu+\frac{\hat{\epsilon}}{2}\epsilon^{\mu\nu\lambda\rho}\partial_\mu G_{\lambda\rho}\ .\lb{EOMF_nonreg_J}
\eea
We note that as opposed to the equations of motion~(\ref{EOMFG_nonreg}), in the absence of sources, these equations are no longer Hodge conjugated to
each other. Then, in order to obtain effective electric and magnetic theories, one has to choose which one to solve.
In order to integrate (hence lower the order) of these equation we are re-writing the action in terms of gauge invariant
quantities~(Schwinger~1951). So let us consider an Hodge decomposition for the currents that obey the continuity equations $\partial_\mu J_e^\mu=\partial_\mu J_g^\mu=0$ required for gauge invariance of the full action,
\be
\ba{rclcrcl}
J_e^\mu&=&\displaystyle\frac{1}{2}\epsilon^{\mu\nu\delta\rho}\partial_\nu\phi^e_{\delta\rho}+c_e^\mu&\ \ ,\ \ \ \ &\partial_\mu\phi^{e\,\mu\nu}&=&0\ ,\\[5mm]
J_g^\mu&=&\displaystyle\frac{1}{2}\epsilon^{\mu\nu\delta\rho}\partial_\nu\phi^g_{\delta\rho}+c_g^\mu&\ \ ,\ \ \ \ &\partial_\mu\phi^{g\,\mu\nu}&=&0\ .
\ea
\lb{HH_J}
\ee
Here $\phi_e$ and $\phi_g$ are close 2-forms and $c_e$ and $c_g$ are constant background currents. In the following we address
local current densities only, hence we set $c_e=c_g=0$. For constant backgrounds due either to topological non-trivial charge
configurations, large gauge transformations or Wilson lines we generally have $c_e\neq 0$ and $c_g\neq 0$. See, for example~(Schwinger~1962, Anderson~1963, Proca~1988, Castelo Ferreira \& Mendon\c{c}a~2006, Mendon\c{c}a \& Castelo Ferreira~2006), where a Proca mass for the surviving field is generated in presence
of background currents.
Using the above Hodge decompositions~(\ref{HH_J}) we can rewrite the action terms containing the current densities
in~(\ref{S_Unbroken_J}) as
\be
S_{\mathrm{Sources}}=-\int_M\left[\left(\frac{1}{4}\epsilon^{\mu\nu\delta\rho}F_{\mu\nu}\phi^e_{\delta\rho}+\frac{\hat{\epsilon}}{2}G^{\mu\nu}\phi^e_{\mu\nu}\right)-\left(\frac{\hat{\epsilon}}{4}\epsilon^{\mu\nu\delta\rho}G_{\mu\nu}\phi^g_{\delta\rho}-\frac{1}{2}F^{\mu\nu}\phi^g_{\mu\nu}\right)\right]\ .
\lb{S_Sources}
\ee
In deriving this expression we have performed an integration by parts and discarded boundary terms. This action is explicitly gauge invariant,
it only depends on the gauge connections, therefore the equations of motion can be written in terms of the gauge connections instead of
its derivatives. For completeness in our discussion, it is also interesting to note that at classical level one can consider the degrees
of freedom to be the gauge connections instead of the gauge fields. This means that, up to integration constants, varying the action with
respect to the gauge fields $A$ and $C$ or the respective gauge connections $F$ and $G$ hold the same equations and we could
consider an effective non-gauge theory. We do not develop this possible construction here, instead we are assuming that the physical
degrees of freedom are encoded in the gauge fields $A$ and $C$.

\subsection{Electrical Solutions for Non-Regular Fields}

Using the Hodge decomposition~(\ref{HH_J}) for the currents, we can integrate equation~(\ref{EOMG_nonreg_J}) obtaining
\be
G^{\mu\nu}=\frac{\hat{\epsilon}}{2}\epsilon^{\mu\nu\lambda\rho}\phi^g_{\lambda\rho}-\frac{\hat{\epsilon}}{2}\epsilon^{\mu\nu\lambda\rho}F_{\lambda\rho}+\frac{\alpha_e}{2}\epsilon^{\mu\nu\lambda\rho}f_{\lambda\rho}\ .
\lb{G_sols_J}
\ee
Here $f$ is defined in terms of a regular field $a$, as given in equation~(\ref{G_nonreg_0}) and, again for convenience, we consider a
constant $\alpha_e=\pm 1$. Considering solution~(\ref{G_sols_J}) and the Hodge decomposition~(\ref{HH_J}) for the currents,
we obtain for each of the terms constituting the action $S=S_0+S_{\mathrm{Sources}}$
given by~(\ref{S_Unbroken_0}) and~(\ref{S_Sources}), the following expressions
\be
\ba{rcl}
\displaystyle+\frac{1}{4}F_{\mu\nu}F^{\mu\nu}&=&\displaystyle+\frac{1}{4}F_{\mu\nu}F^{\mu\nu}\ ,\\[2mm]
\displaystyle-\frac{1}{4}G_{\mu\nu}G^{\mu\nu}&=&\displaystyle+\frac{1}{4}F_{\mu\nu}F^{\mu\nu}+\frac{\alpha_e^2}{4}f_{\mu\nu}f^{\mu\nu}-\frac{2\hat{\epsilon}\alpha_e}{4}F_{\mu\nu}f^{\mu\nu}\\[2mm]
                      & &\displaystyle-\frac{2}{4}F^{\mu\nu}\phi^g_{\mu\nu}+\frac{2\hat{\epsilon}\alpha_e}{4}f^{\mu\nu}\phi^g_{\mu\nu}+\frac{1}{4}\phi^g_{\mu\nu}\phi^{g\,\mu\nu}\ ,\\[2mm]
\displaystyle-\frac{\hat{\epsilon}}{4}\epsilon^{\mu\nu\lambda\rho}F_{\mu\nu}G_{\lambda\rho}&=&\displaystyle-\frac{2}{4}F_{\mu\nu}F^{\mu\nu}+\frac{2\hat{\epsilon}\alpha_e}{4}F_{\mu\nu}f^{\mu\nu}+\frac{2}{4}F^{\mu\nu}\phi^g_{\mu\nu}\ ,\\[2mm]
\displaystyle+\frac{\hat{\epsilon}}{2}G^{\mu\nu}\phi^e_{\mu\nu}&=&\displaystyle+\frac{1}{4}\epsilon^{\mu\nu\lambda\rho}\phi^e_{\mu\nu}\phi^g_{\lambda\rho}-\frac{1}{4}\epsilon^{\mu\nu\lambda\rho}F_{\mu\nu}\phi^e_{\lambda\rho}+\frac{\hat{\epsilon}\alpha_e}{4}\epsilon^{\mu\nu\lambda\rho}f_{\mu\nu}\phi^e_{\lambda\rho}\ ,\\[2mm]
\displaystyle-\frac{\hat{\epsilon}}{4}\epsilon^{\mu\nu\lambda\rho}G_{\mu\nu}\phi^g_{\lambda\rho}&=&\displaystyle+\frac{2}{4}\phi^g_{\mu\nu}\phi^{g\,\mu\nu}-\frac{2}{4}F^{\mu\nu}\phi^g_{\mu\nu}+\frac{2\hat{\epsilon}\alpha_e}{4}f^{\mu\nu}\phi^g_{\mu\nu}\ ,\\[2mm]
\displaystyle+\frac{1}{4}\epsilon^{\mu\nu\lambda\rho}F_{\mu\nu}\phi^e_{\lambda\rho}&=&\displaystyle+\frac{1}{4}\epsilon^{\mu\nu\lambda\rho}F_{\mu\nu}\phi^e_{\lambda\rho}\ ,\\[2mm]
\displaystyle+\frac{1}{2}F^{\mu\nu}\phi^g_{\mu\nu}&=&\displaystyle+\frac{1}{2}F^{\mu\nu}\phi^g_{\mu\nu}\ .
\ea
\lb{terms_e_nonreg_J}
\ee
Where, again, we used identity~(\ref{HH_id}). Hence the effective electric action is
\be
\ba{rcl}
S_{\mathrm{Electric}}&=&\displaystyle-\int_M\left[\frac{1}{4}f_{\mu\nu}f^{\mu\nu}+\frac{\hat{\epsilon}\alpha_e}{4}\epsilon^{\mu\nu\delta\rho}f_{\mu\nu}\phi^e_{\delta\rho}+\hat{\epsilon}\alpha_ef^{\mu\nu}\phi^g_{\mu\nu}\right.\\[2mm]
                     & &\ \ \ \ \ \ \ \ \ \ \displaystyle\left.+\frac{3}{4}\phi^{g\,\mu\nu}\phi^g_{\mu\nu}+\frac{1}{4}\epsilon^{\mu\nu\delta\rho}\phi^e_{\mu\nu}\phi^g_{\delta\rho}\right]\ .
\ea
\lb{el_eff_J}
\ee
The last two terms contribute to the vacuum energy and, at classical level, as long as we are not dealing with gravity, are irrelevant.
For $\hat{\epsilon}\alpha_e=+1$, the first three terms, are recognized as the Maxwell action in the presence of
local electric and magnetic charges~(Schwinger~1951). Also it is important to stress that for regular $a$ and
the currents obeying the Hodge decomposition~(\ref{HH_J}), the third term is a total derivative. Then
we obtain the effective action (up to a vacuum energy shift) and the effective electromagnetic fields definitions
\be
\ba{l}
\displaystyle S_{\mathrm{Electric}}=-\int_M\left[\frac{1}{4}f_{\mu\nu}f^{\mu\nu}+a_\mu J_e^\mu\right]\ ,\\[2mm]
\displaystyle E^i=\hat{\epsilon}\alpha_ef^{0i}+\phi^{g\,0i}\ \ ,\ \  
B^i=\frac{\hat{\epsilon}\alpha_e}{2}\epsilon^{0ijk}f_{jk}+\frac{1}{2}\epsilon^{0ijk}\phi^g_{jk}\ .
\ea
\lb{el_eff_0_J}
\ee
In deriving these results we have integrated by parts the second and third terms of~(\ref{el_eff_J}) and
considered the identity $\epsilon^{0ijk}\epsilon_{jk}^{\ \ \,0i'}\phi^g_{0i'}=-2\phi^{g\,0i}$.
This is the standard Maxwell action in the presence of electric sources and the field expressions
are obtained by the direct replacement of the solution~(\ref{G_sols_J}) for $G$ in the fields definitions~(\ref{ph_fields_FG})
of the unbroken theory. For $\hat{\epsilon}\alpha_e=+1$ these expressions correspond to the standard definitions in electromagnetism
plus a contribution from the magnetic sources. This means that the magnetic current effects are still present in the effective theory,
not at the level of the action but at the level of the electromagnetic fields definitions which are inherit from the unbroken theory.

\subsection{Magnetic Solutions for Non-Regular Fields}

Using the Hodge decomposition~(\ref{HH_J}) for the currents we obtain that the generic solution for the equations
of motion~(\ref{EOMF_nonreg_J}) is
\be
F^{\mu\nu}=\frac{1}{2}\epsilon^{\mu\nu\lambda\rho}\phi^e_{\lambda\rho}+\frac{\hat{\epsilon}}{2}\epsilon^{\mu\nu\lambda\rho}G_{\lambda\rho}+\frac{\alpha_g}{2}\epsilon^{\mu\nu\lambda\rho}g_{\lambda\rho}\ .
\lb{F_sols_J}
\ee
Again the field $g$ is defined in terms of a regular field $c$, as given in equation~(\ref{F_nonreg_0}), and once more for convenience, we consider a
constant $\alpha_g=\pm 1$. Replacing the solution~(\ref{F_sols_J}) in the action $S=S_0+S_{\mathrm{Sources}}$
given by~(\ref{S_Unbroken_0}) and~(\ref{S_Sources}) we obtain for each of the terms constituting the action the following expressions
\be
\ba{rcl}
\displaystyle-\frac{1}{4}G_{\mu\nu}G^{\mu\nu}&=&\displaystyle-\frac{1}{4}G_{\mu\nu}G^{\mu\nu}\ ,\\[2mm]
\displaystyle+\frac{1}{4}F_{\mu\nu}F^{\mu\nu}&=&\displaystyle-\frac{1}{4}G_{\mu\nu}G^{\mu\nu}-\frac{\alpha_g}{4}g_{\mu\nu}g^{\mu\nu}-\frac{2\hat{\epsilon}\alpha_g}{4}G_{\mu\nu}g^{\mu\nu}\\[2mm]
                                             & &\displaystyle-\frac{2\hat{\epsilon}}{4}G^{\mu\nu}\phi^e_{\mu\nu}-\frac{2\alpha_g}{4}g^{\mu\nu}\phi^e_{\mu\nu}-\frac{1}{4}
\phi^e_{\mu\nu}\phi^{e\,\mu\nu}\ ,\\[2mm]
\displaystyle-\frac{\hat{\epsilon}}{4}\epsilon^{\mu\nu\lambda\rho}F_{\mu\nu}G_{\lambda\rho}&=&\displaystyle+\frac{2}{4}G_{\mu\nu}G^{\mu\nu}+\frac{2}{4}\hat{\epsilon}\alpha_gG_{\mu\nu}g^{\mu\nu}+\frac{2\hat{\epsilon}}{4}G^{\mu\nu}\phi^e_{\mu\nu}\ ,\\[2mm]
\displaystyle+\frac{1}{4}\epsilon^{\mu\nu\lambda\rho}F_{\mu\nu}\phi^e_{\lambda\rho}&=&\displaystyle-\frac{2}{4}\phi^e_{\mu\nu}\phi^{e\,\mu\nu}-\frac{2\hat{\epsilon}}{4}G^{\mu\nu}\phi^e_{\mu\nu}-\frac{2\alpha_g}{4}g^{\mu\nu}\phi^{e}_{\mu\nu}\ ,\\[2mm]
\displaystyle+\frac{\hat{\epsilon}}{2}G^{\mu\nu}\phi^e_{\mu\nu}&=&\displaystyle+\frac{\hat{\epsilon}}{2}G^{\mu\nu}\phi^e_{\mu\nu}\ ,\\[2mm]
\displaystyle-\frac{\hat{\epsilon}}{4}\epsilon^{\mu\nu\lambda\rho}G_{\mu\nu}\phi^g_{\lambda\rho}&=&\displaystyle-\frac{\hat{\epsilon}}{4}\epsilon^{\mu\nu\lambda\rho}G_{\mu\nu}\phi^g_{\lambda\rho}\ ,\\[2mm]
\displaystyle+\frac{1}{2}F^{\mu\nu}\phi^g_{\mu\nu}&=&\displaystyle+\frac{1}{4}\epsilon^{\mu\nu\lambda\rho}\phi^e_{\mu\nu}\phi^g_{\lambda\rho}+\frac{\hat{\epsilon}}{4}\epsilon^{\mu\nu\lambda\rho}G_{\mu\nu}\phi^g_{\lambda\rho}+\frac{\alpha_g}{4}\epsilon^{\mu\nu\lambda\rho}g_{\mu\nu}\phi^g_{\lambda\rho}\ .
\ea
\lb{terms_g_nonreg_J}
\ee
Where again we used the identity~(\ref{HH_id}). Hence we obtain the effective magnetic action
\be
\ba{rcl}
S_{\mathrm{Magnetic}}&=&\displaystyle-\int_M\left[-\frac{1}{4}g_{\mu\nu}g^{\mu\nu}+\frac{\alpha_g}{4}\epsilon^{\mu\nu\lambda\rho}g_{\mu\nu}\phi^g_{\lambda\rho}-\alpha_gg^{\mu\nu}\phi^e_{\mu\nu}\right.\\[2mm]
                     & &\ \ \ \ \ \ \ \ \ \displaystyle\left.-\frac{3}{4}\phi^e_{\mu\nu}\phi^{e\,\mu\nu}+\frac{1}{4}\epsilon^{\mu\nu\lambda\rho}\phi^e_{\mu\nu}\phi^g_{\lambda\rho}\right]\ .
\ea
\lb{mg_eff_J}
\ee
As in the electric case, the last two terms contribute to the vacuum energy. For $\alpha_g=-\hat{\epsilon}$, the first three terms
are recognized as the magnetic counterpart of the Maxwell action in the presence of local electric and magnetic charges~(Castelo Ferreira~2006).
For $c$ regular and the currents obeying the Hodge decomposition~(\ref{HH_J}) the third term is a total derivative. Then we obtain
the effective action and the electromagnetic field definitions
\be
\ba{l}
\displaystyle S_{\mathrm{Magnetic}}=\int_M\left[\frac{1}{4}g_{\mu\nu}g^{\mu\nu}+\hat{\epsilon}c_\mu J_g^\mu\right]\\[2mm]
\displaystyle E^i=-\frac{\hat{\epsilon}}{2}\epsilon^{0ijk}g_{jk}+\frac{1}{2}\epsilon^{0ijk}\phi^e_{jk}\ \ ,\ \ 
B^i=+\hat{\epsilon}g^{0i}-\phi^{e\,0i}\ .
\ea
\lb{mg_eff_0_J}
\ee
This action is the magnetic counterpart of the Maxwell action in the presence of magnetic sources and
the effective field definitions have, in this case, a correction due to the presence of electric sources.

\subsection{Solutions for Regular Fields}

For regular gauge fields the equations of motion for the action $S=S_0+S_{\mathrm{Sources}}$
given by~(\ref{S_Unbroken_0}) and~(\ref{S_Unbroken_J}) are obtained by a variation with respect to $C$ and $A$, holding
the equations
\be
\partial_\mu G^{\mu\nu}=\hat{\epsilon}J_g^\nu\ \ ,\ \ \partial_\mu F^{\mu\nu}=J_e^\nu\ .\lb{EOMFG_reg_J}
\ee
Using the current decompositions~(\ref{HH_J}) the generic solutions for these equations are, respectively
\be
G^{\mu\nu}=\frac{\hat{\epsilon}}{2}\epsilon^{\mu\nu\delta\rho}\phi^g_{\delta\rho}+H_g^{\mu\nu}\ \ ,\ \ F^{\mu\nu}=\frac{1}{2}\epsilon^{\mu\nu\delta\rho}\phi^e_{\delta\rho}+H_e^{\mu\nu}\ ,\lb{EOM_FG_J}
\ee
where $H_e$ and $H_g$ are closed 2-forms such that $\partial_\mu H_e^{\mu\nu}=\partial_\mu H_g^{\mu\nu}=0$.
As we did in section~\ref{sols} we can consider a field redefinition of the form
\be
H_g^{\mu\nu}=-\frac{\hat{\epsilon}}{2}\epsilon^{\mu\nu\lambda\rho}F^{\lambda\rho}+\frac{\alpha_e}{2}\epsilon^{\mu\nu\lambda\rho}f^{\lambda\rho}\ \  ,\ \ H_e^{\mu\nu}=\frac{\hat{\epsilon}}{2}\epsilon^{\mu\nu\lambda\rho}G^{\delta\rho}+\frac{\alpha_g}{2}\epsilon^{\mu\nu\lambda\rho}g^{\lambda\rho}\ ,
\ee
such that we obtain the same expressions for the solutions of non-regular gauge fields, as given in~(\ref{G_sols_J}) and~(\ref{F_sols_J}).
The remaining of the proof follows in the same manner as for non-regular gauge fields, such that we obtain the same effective actions
and electromagnetic field redefinitions given in~(\ref{el_eff_0_J}) and~(\ref{mg_eff_0_J}). Again we stress that the coupling
between both sectors is only fully justified by considering compatibility with the existence of non-regular field configurations.

\subsection{\lb{quant}Maxwell Equations and Current Densities Definitions}

Let us address what are the consequences of the above results at the level of the Maxwell equations,
in particular of the physical fields definitions~(\ref{el_eff_0_J}) in the effective theory that, as we have seen,
are inherit from the physical field definitions~(\ref{ph_fields_FG}) of the original unbroken theory.
In the electric broken theory the equations of motion and the Bianchi identities are given, respectively, by
\be
\partial_\mu f^{\mu\nu}= J_e^\nu\ \ ,\ \ \epsilon^{\nu\mu\lambda\rho}\partial_\mu f_{\lambda\rho}=0\ .
\lb{EOM_f_J}
\ee
The equations of motion are obtained consistently either by direct replacement of the solution~(\ref{G_sols_J}) in the equation
of motion~(\ref{EOMFG_nonreg}) or by a variation of the effective action~(\ref{el_eff_0_J}) with respect to the field $a_\nu$ while
the Bianchi identities are due to the field $a$ being regular by construction as given in~(\ref{G_sols_J}) and~(\ref{G_nonreg_0}).
Equations~(\ref{EOM_f_J}) correspond to the 8 Maxwell equations ($\nu=0,1,2,3$), in order to rewrite this
equations in terms of the electric and magnetic fields definitions~(\ref{el_eff_0_J}) let us define the tensor ${\mathcal{F}}$
\be
{\mathcal{F}}^{0i}= E^i\ \ ,\ \ {\mathcal{F}}^{ij}=-\frac{1}{2}\epsilon^{ijk}B^k\ \ ,\ \ 
f_{\mu\nu}={\mathcal{F}}_{\mu\nu}-\phi^g_{\mu\nu}\ .
\lb{FF_f}
\ee
Hence the Maxwell equations are written as
\be
\partial_\mu {\mathcal{F}}^{\mu\nu}= J_e^\nu\ \ ,\ \ \epsilon^{\nu\mu\lambda\rho}\partial_\mu {\mathcal{F}}_{\lambda\rho}=J_g^\nu\ ,
\lb{Maxwell_f_J}
\ee
which are straight forwardly recognized as the generalized standard Maxwell equations in the presence of magnetic currents~(Jackson~1975).
In the first equation the term dependent on the magnetic current is null according to the Hodge decomposition~(\ref{HH_J}),
i.e. $\partial_\mu\phi^{g\,\mu\nu}=0$. In the second equation we have used the Hodge decomposition~(\ref{HH_J})
in order to rewrite the equation in terms of the magnetic current $J_g^\mu=\epsilon^{\mu\nu\delta\rho}\partial_\nu\phi^g_{\delta\rho}/2$.
So we manage to recover the effects of magnetic currents in the broken theory at the level of the 
Maxwell equations. The gauge fields are regular and the re-appearance of magnetic current is due to the electric and magnetic field
definitions~(\ref{el_eff_0_J}). Although this construction may seem to consist in a simple redefinition of the gauge
field $a_\mu$, it is not so. In order to see it explicitly, let us note that $\phi^g_{\mu\nu}$ could be written in terms of
a four-vector field $\varphi^g_\mu$, obeying the following equations
\be
\phi^g_{\mu\nu}=\partial_\mu\varphi^g_\nu-\partial_\nu\varphi^g_\mu\ \ ,\ \ \epsilon^{\mu\nu\lambda\rho}\partial^g_\nu\partial_\lambda\varphi^g_\rho=J_g^\mu\ \ ,\ \ \partial_\nu\partial^\nu\varphi^{g\,\mu}=\partial_\nu\partial^\mu\varphi^{g\,\nu}\ .
\lb{varphi}
\ee
The second and third equations are obtained directly from the Hodge decomposition~(\ref{HH_J}) and are
due to requiring compatible with the continuity equation $\partial_\mu J_g^\mu$ (hence gauge invariance).
We readily conclude that indeed, the dependence on $\phi^g_{\mu\nu}$ of the field definitions~(\ref{el_eff_0_J}),
cannot possible be eliminated by a redefinition of the gauge field $a_\mu$, or equivalently of the gauge connection $f_{\mu\nu}$.
By construction $a_\mu$ is a regular field obeying the Bianchi identities ($\epsilon^{\mu\nu\lambda\rho}\partial_\nu f_{\lambda\rho}=0$),
while the field $\varphi^g_\mu$ is non-regular, its Bianchi identity is violated as given by the second equation in~(\ref{varphi}).
Given these results, and for completeness, we also note that $f_{\mu\nu}$ and $\phi^g_{\mu\nu}$ correspond,
generally, to the two components of an Hodge decomposition of the tensor ${\mathcal{F}}_{\mu\nu}$ defined
in equation~(\ref{FF_f}), that can also be defined in terms of a general field ${\mathcal{A}}_\mu=a_\mu+\varphi_\mu^g$ such that
its regular and non-regular parts are respectively $a_\mu$ and $\varphi^g_\mu$. If one considers this field to be the physical gauge field
we would simply obtain the original singularities in the gauge fields (the Dirac string and Wu-Yang fiber-bundle).
Let us stress the crucial differences between our approach and the approaches which consider
the gauge field ${\mathcal{A}}$ as the physical field: In the present construction we simply are not allowed
to redefine the physical field $a_\mu\to a_\mu+\varphi_\mu$, by construction this field is regular.
There is also another important point concerning our last comment, the full unbroken theory
contains both photons and pseudo-photons degrees of freedom, while in any of the effective broken theories
half of the degrees of freedom are truncated, nevertheless the physical field content in the broken theories
is inherited from the unbroken theory and we cannot freely choose what our fields are, they are constraint by the
field configurations used in the dynamical symmetry breaking. Moreover, by the end of this section, we will have prove
non-triviality of the full unbroken theory in the presence of both electric and magnetic monopoles which implies that it is not possible to
describe the theory by the standard $U(1)$ variational Maxwell theory. These crucial remarks
clearly distinguish our broken effective theories from standard variational electromagnetism in the presence of magnetic monopoles.
By construction the singularities are not included, and cannot be included, in the gauge fields.

In the effective theory the electric and magnetic charge ($Q_e$ and $Q_g$) and the electric and magnetic 3-current
fluxes ($j^i_{\phi\,e}$ and $j^i_{\phi\,g}$) are defined as
\be
\ba{rcl}
Q_e&=&\displaystyle \int_M dx^3\partial_if^{0i}+\frac{1}{2}\int_{M}dx^3\partial_i\phi^{g\,0i}=\oint_{\partial M}dx^2f^{0i}n_i=\int_M dx^3 J^0_e\ ,\\[2mm]
j^i_{\phi\,e}&=&\displaystyle \int_M dx^3\partial_jf^{ji}+\frac{1}{2}\int_{M}dx^3\partial_j\phi^{g\,ji}=\oint_{\partial M}dx^2f^{ij}n_i=\int_M dx^3 J^i_e\ ,\\[2mm]
Q_g&=&\displaystyle \int_M dx^3\epsilon^{ijk}\partial_i f_{jk}+\frac{1}{2}\int_{M}dx^3\epsilon^{ijk}\partial_i\phi^g_{jk}=\frac{1}{2}\oint_{\partial M}dx^2\epsilon^{ijk}\phi^g_{jk}n_i=\int_M dx^3 J^0_g\ ,\\[2mm]
j_{\phi\,g}^i&=&\displaystyle \int_M dx^3\left[\epsilon^{ijk}\partial_0 f_{jk}-2\epsilon^{ijk}\partial_j f_{0k}\right]+\frac{1}{2}\int_{M}dx^3\left[\epsilon^{ijk}\partial_0\phi^g_{jk}-2\epsilon^{ijk}\partial_j\phi^g_{0k}\right]\\[2mm]
&=&\displaystyle\frac{1}{2}\int_{M}dx^3\epsilon^{ijk}\dot{\phi}^g_{jk}-\oint_{\partial M}dx^2\epsilon^{ijk}\phi^g_{0k}n_j=\int_M dx^3 J^i_g\ .
\ea
\lb{Q_EM}
\ee
This result is consistent with the previous results. Electric charge corresponds, as usual,
to the $U(1)$ group charge, however we must stress that magnetic charge definition is inherit from the
unbroken theory and cannot, in the broken theory, be interpreted as group topological charge.
Nevertheless the canonical variables of the theory are going to be sensitive to these currents.
The canonical momenta of the unbroken theory and the respective effective canonical momenta for the broken electric theory are
\be
\pi_A^i=F^{0i}-\frac{\hat{\epsilon}}{2}\epsilon^{0ijk}G_{jk}=E^i\ \ \Rightarrow\ \ \pi_A^i=\hat{\epsilon}\alpha_ef^{0i}+\phi^{g\,0i}=E^i\ .
\lb{pi_A}
\ee
The first expression is obtained by varying the action~(\ref{S_Unbroken_0}) with respect to $\delta \partial_0 A_i$~(Castelo Ferreira~2006)
and the second expression is obtained by replacing the solution~(\ref{G_sols_J}) for the $C$ field in the first expression.
This is the same procedure we employed to obtain the effective definitions~(\ref{el_eff_0_J}) of electromagnetic fields. For the case of
of canonical momenta we can further derive this result directly from the broken theory. By noting that we have integrated
by parts the third term in action~(\ref{el_eff_J}) we obtain a \textit{time-boundary} term of the form
\be
S_{\mathrm{bound}}=\hat{\epsilon}\alpha_e\left.\int d^3x A_i\phi^{g\,0i}\right|_{t_1}^{t_2}\ .
\ee
This is actually not a standard boundary term, in more exact terms are initial conditions for the equations of motion
that ensure global conservation of magnetic charge. Nevertheless its effect holds the correction~(\ref{pi_A}),
as long as $\hat{\epsilon}\alpha_e=+1$, which coincides with the choice corresponding to the standard definition
of electromagnetic fields in~(\ref{el_eff_0_J}). Similar results hold for the magnetic effective theory.

As a very simple example let us consider the effective electric theory of a point-like static electric charge $q_e$ located at
$\vb{r_e}=(x^1_e,x^2_e,x^3_e)$ and a point-like magnetic charge $q_g$ located at $\vb{r_g}=(x^1_g,x^2_g,x^3_g)$.
The solution for the $a$ field at a given point $\vb{r}=(x^1,x^2,x^3)$ is the standard one, while the solution for $\phi_{\mu\nu}$
can be inferred from the solution for $G_{\mu\nu}$ corresponding to a magnetic monopole in the unbroken theory, i.e.
\be
a_0(\vb{r})=\frac{q_e}{|\vb{r}-\vb{r_e}|}\ \ ,\ \ a_i(\vb{r})=0\ \ ,\ \ \phi^g_{0i}=0\ \ ,\ \ \phi^g_{ij}=\frac{q_g\epsilon_{ijk}(x^k-x_e^k)}{|\vb{r}-\vb{r_g}|^3}\ .
\ee
As usual, due to considering point-like charges, we have singularities at $\vb{r}=\vb{r}_e$ and $\vb{r}=\vb{r}_g$,
however as expected we have no extended singularities in the gauge field $a$ of $\phi$.
The electromagnetic fields definitions~(\ref{el_eff_0_J}) and the space-time equations of motion (Lorentz force)
for a particle of electric charge $q$ and mass $m$ are
\be
\ba{l}
\displaystyle E^i(\vb{r})=\frac{q_e (x^i-x_e^i)}{|\vb{r}-\vb{r_e}|^3}\ \ ,\ \ B^i(\vb{r})=\frac{q_g(x^i-x_g^i)}{|\vb{r}-\vb{r_g}|^3}\\[3mm]
\displaystyle m\ddot{x}^i=\frac{q\,q_e (x^i-x_e^i)}{|\vb{r}-\vb{r_e}|^3}+\frac{q\,q_g\epsilon^{i}_ {\ \,jk}\,\dot{x}^j(x^k-x_g^k)}{|\vb{r}-\vb{r_g}|^3}\ .
\ea
\ee
Accordingly to the charge definitions~(\ref{Q_EM}) the total electric charge is $Q_e=q_e$ and the total magnetic charge is $Q_g=q_g$.

\subsection{Again on Trivial Solutions}

The trivial solutions~(\ref{sols_trivial}) are no longer valid, we note that by including the source terms
the equations of motion for $G$~(\ref{EOMG_nonreg_J}) and $F$~(\ref{EOMF_nonreg_J}) are no longer Hodge conjugate
to each other, hence are linear independent (as opposed to~(\ref{EOMFG_nonreg}) in the absence of sources).
Also, in the presence of generic four-currents, considering the trivial solutions corresponding
to solutions~(\ref{G_sols_J}) and~(\ref{F_sols_J}) with $\alpha_e=0$ and $\alpha_g=0$ is not possible.
Simply they are not compatible with each other. Let us exemplify it, take the case of the electric trivial solution
\be
G^{\mu\nu}=\frac{\hat{\epsilon}}{2}\epsilon^{\mu\nu\lambda\rho}\phi^g_{\lambda\rho}-\frac{\hat{\epsilon}}{2}\epsilon^{\mu\nu\lambda\rho}F_{\lambda\rho}\ \ \Rightarrow\ \left\{\ba{l} \displaystyle J_e^\nu=\hat{\epsilon}\partial_\mu\phi^{g\,\mu\nu}=0\\[2mm]
\displaystyle\frac{\alpha_g}{2}\epsilon^{\mu\nu\lambda\rho}g_{\lambda\rho}=0\ ,\ea\right.
\ee
where the first expression corresponds to solution~(\ref{G_sols_J}) with $\alpha_e=0$ (which is a solution of the equation of motion~(\ref{EOMG_nonreg_J})) and the second ones to the remaining equation of motion~(\ref{EOMF_nonreg_J}) corresponding to
this particular solution. Due to $g_{\lambda\rho}$ being regular by construction, this equation decouples in its Hodge components
which must, independently of each other, obey the equation. In the presence of electric and magnetic charge these are not obeyed,
hence the trivial solutions are not extrema of the action and the only allowed solutions are actually of the form~(\ref{G_sols_J})
and~(\ref{F_sols_J}) with $\alpha_e\neq 0$ and $\alpha_e\neq 0$. The respective broken theories have, for these non-trivial solutions,
consistent equations of motion, as given in~(\ref{Maxwell_f_J})
for the electric case. It is interesting to note that this result also shows non-triviality of the unbroken theory.
The same arguments are valid if, instead of magnetic monopoles, one considers external non-regular fields as in~(Castelo Ferreira~2006).
In addition these results also explain why the broken effective theories are not
equivalent to standard electromagnetism, as already mentioned before, the \textit{broken} physical gauge fields cannot
be freely chosen, they must be compatible with the unbroken theory, in particular are regular. Hence we have conclude our
prove of non-equivalence between extended $U_e(1)\times U_g(1)$ electromagnetism and standard $U(1)$ variational Maxwell theory.
However we stress that our construction is in complete agreement with the Maxwell equations which, we recall, have been
inferred phenomenologically from experiments.

\section{\lb{dof}Singularities, Gauge Invariance and Zero-Field Equation}

\subsection{Singularities and Gauge Symmetry Breaking}

Let us resume the construction developed in the previous sections. By solving the equations of motion
of the original theory containing two gauge fields we have managed to obtain one broken
electric and one broken magnetic effective theories with only one gauge field. We achieve these by choosing to replace in the original
action either $G$ or $F$ such that, at functional level, we are expressing each gauge field as a functional of the
other one, respectively $C=C(A)$ and $A=A(C)$. In our construction the effective theories given by actions~(\ref{el_eff_0_J})
and~(\ref{mg_eff_0_J}) are defined in terms of only regular fields $a$ and $c$. Hence the usual singularities
and/or discontinuities of the fields~(Dirac,Wu \& Yang~1975) are encoded in the non-trivial solutions
for the effectively excluded fields. This is the main novelty in our construction and we note that a similar feature is
already considered in~(Cardoso de Mello \& al.~1996, Berkovits~1996, Carneiro~1997). Concerning the interpretation in terms of the degrees of freedom of the theory
we note the dynamical symmetry breaking conditions consist of 4~independent equations that constraint 8 field components
$A_\mu$ and $C_\mu$ ($\mu=0,1,2,3$). Therefore as expected we obtain 4 independent gauge fields encoded in $a_\mu$ or
$c_\mu$, we start with 6 degrees of freedom (4 physical) corresponding to one massless photon and one massless
pseudo-photon and end up with only 3 degrees of freedom (2 physical) corresponding to either the massless photon
or the massless pseudo-photon. As for gauge invariance there is a subtlety in the gauge symmetry breaking mechanism.
We start by having a gauge symmetry $U_e(1)\times U_g(1)$ and broke it down to one single $U(1)$ (either $U_{\bar{e}}(1)$
or $U_{\bar{g}}(1)$), however we did not properly address what is the exact relation between the surviving symmetry and
the original gauge symmetry. In order to see it explicitly take for example solution~(\ref{G_nonreg_0}), as it
stands this solution is gauge invariant under any of the groups. However expressing it
in terms of the gauge fields instead of the connections, that is no longer the case
\be
\tilde{C}_\mu=\hat{\epsilon}\,A_\mu-\alpha_e\,a_\mu\ .
\lb{Eq_C}
\ee
This solution is defined up to a closed 1-form uncharged under both groups, although it can be offset by a gauge transformation
it does not change any of the arguments that follow. Let us consider a $U_e(1)\times U_g(1)$ gauge transformations with regular
gauge parameters $\Lambda_e$ and $\Lambda_g$ for some fields $\bar{A}_\mu$, $\bar{C}_\mu$ and $\bar{a}_\mu$
\be
A_\mu=\bar{A}_\mu+\partial_\mu\Lambda_e-\hat{\epsilon}\partial_\mu\tilde{\Lambda}_g\ \ ,\ \ C_\mu=\bar{C}_\mu+\partial_\mu\Lambda_g+\hat{\epsilon}\partial_\mu\tilde{\Lambda}_e\ .
\lb{dLambda}
\ee
Each field is charged under both $U_e(1)$ and $U_g(1)$ as can explicitly be seen from
the current couplings~(\ref{S_Unbroken_J}), however we note that each of the fields is charged under one
of the group currents and is topological charged with respect to the other group~(Castelo Ferreira~2006).
Considering these gauge transformations~(\ref{dLambda}) in the solution for the field $C$~(\ref{Eq_C}) we obtain
\be
\tilde{\bar{C}}_\mu=\hat{\epsilon}\,\bar{A}_\mu-\alpha_e\,\bar{a}_\mu-2\partial_\mu\tilde{\Lambda}_g+2\hat{\epsilon}\partial_\mu\Lambda_e=\hat{\epsilon}\,\bar{A}_\mu-\alpha_e\,\bar{a}_\mu+2\hat{\epsilon}\partial_\mu\Lambda_e\ .
\lb{gauge_trans}
\ee
As long as the parameter $\Lambda_g$ is regular $\tilde{\Lambda}_g=0$, as expressed by
the dual fields definition~(\ref{dual_fields}). Then the equivalent parameter of the gauge transformation
corresponding to the field $a$ that reproduces the gauge transformation in~(\ref{gauge_trans}) is
\be
\Lambda_{\bar{e}}=-\frac{2\hat{\epsilon}}{\alpha_e}\Lambda_e\ .
\lb{map_e}
\ee
This equation constitutes a map between the original $U_e(1)$ and $U_{\bar{e}}(1)$ such that $U_{\bar{e}}(1)\cong U_e(1)$.
For the standard integration constants choice~(\ref{signs_0}) we have $\alpha_e=\hat{\epsilon}$ and this map corresponds both,
to a sign inversion, and a rescaling of the gauge parameter. Therefore we conclude that the gauge symmetry corresponding to
the surviving field $a$, in the effective electric theory, is $U_{\bar{e}}(1)$ corresponding to the symmetry breaking
$U_e(1)\times U_g(1)\to U_{\bar{e}}(1)$. For the magnetic effective theory a similar result holds and we obtain the map
$\Lambda_{\bar{g}}=-2\hat{\epsilon}\Lambda_g/\alpha_g$ corresponding to the gauge symmetry $U_e(1)\times U_g(1)\to U_{\bar{g}}(1)$.

\subsection{Hodge Duality as a Symmetry and Zero-Field Equation}

Next we discuss a duality that constitutes an explicit symmetry of the action and its relation with the zero-field
equation~(\ref{0-field}). This condition was considered in the original studies on magnetic monopoles~(Dirac, 1948,Cabibbo \& Ferrari~1962) and
our main motivation is to investigate if it has some special meaning in relation to our mechanism
of dynamical symmetry breaking. Let us start by noting that, relating each of the gauge connection
(or equivalently each gauge field) trough Hodge duality, leaves the action invariant. Considering
the duality transformations
\be
\left\{\ba{rcl}\displaystyle F_{\mu\nu}&\to&\displaystyle-\frac{\hat{\epsilon}}{2}\epsilon^{\mu\nu\lambda\rho}G_{\lambda\rho}\\[2mm]
G_{\mu\nu}&\to&\displaystyle+\frac{\hat{\epsilon}}{2}\epsilon^{\mu\nu\lambda\rho}F_{\lambda\rho}\ea\right.\Leftrightarrow \left\{\ba{rcl}\displaystyle A_{\mu}&\to&\displaystyle-\hat{\epsilon}\,\tilde{C}_{\mu}\\[2mm]
\displaystyle C_{\mu}&\to&\displaystyle+\hat{\epsilon}\,\tilde{A}_{\mu}\ \ ,\ea\right.
\lb{duality}
\ee
we obtain that the several terms in the actions~(\ref{S_Unbroken_0}) and~(\ref{S_Unbroken_J}) transform as
\be
\left\{\ba{rcl}\displaystyle+\frac{1}{4}F_{\mu\nu}F^{\mu\nu}&\to&\displaystyle-\frac{1}{4}G_{\mu\nu}G^{\mu\nu}\\[2mm]
\displaystyle-\frac{1}{4}G_{\mu\nu}G^{\mu\nu}&\to&\displaystyle+\frac{1}{4}F_{\mu\nu}F^{\mu\nu}\\[2mm]
\displaystyle-\frac{\hat{\epsilon}}{4}\epsilon^{\mu\nu\lambda\rho}G_{\mu\nu}F_{\lambda\rho}&\to&\displaystyle-\frac{\hat{\epsilon}}{4}\epsilon^{\mu\nu\lambda\rho}G_{\mu\nu}F_{\lambda\rho}\\[2mm]
\displaystyle+(A_\mu-\hat{\epsilon}\tilde{C}_\mu)J_e^{\mu}&\to&\displaystyle+(A_\mu-\hat{\epsilon}\tilde{C}_\mu)J_e^{\mu}\\[2mm]
\displaystyle-(\hat{\epsilon}C+\tilde{A}_\mu)J_g^{\mu}&\to&\displaystyle-(\hat{\epsilon}C+\tilde{A}_\mu)J_g^{\mu}\ ,\ea\right.
\ee
hence the duality~(\ref{duality}) leaves the Hodge terms invariant and transform the Maxwell terms into each other,
being an exact global symmetry of the action, i.e. $S_0+S_{\mathrm{Sources}}\to S_0+S_{\mathrm{Sources}}$.
Also it is straight forward to show that both the electromagnetic fields~(\ref{ph_fields_FG}) are
invariant under this duality. The most interesting fact is that the zero-field equation~(\ref{0-field})
corresponds to the field self-dual condition $\tilde{C}_\mu=-\hat{\epsilon}A_\mu$ corresponding to the
following relation between the field $a_\mu$ and $A_\mu$,
\be
a_\mu=\frac{2\hat{\epsilon}}{\alpha_e}A^{\mathrm{reg}}_\mu\ \ \Rightarrow\ \ \tilde{C}_\mu=\hat{\epsilon}\,A^{\mathrm{nreg}}_\mu-\hat{\epsilon}\,A^{\mathrm{reg}}_\mu\ .
\lb{0-field_a}
\ee
We have explicitly decomposed the $A=A^{\mathrm{reg}}+A^{\mathrm{nreg}}$ field into a regular part $A^{\mathrm{reg}}$ and
an non-regular part $A^{\mathrm{nreg}}$. This is a generic approach and is due to the field $a$ being regular by construction,
such that the regular parts and the non-regular parts of the equations must be treated independently. Relating this result with the above gauge
symmetry breaking we note that it corresponds to the case for which the map~(\ref{map_e}) does not holds a scaling
of the gauge parameter. Moreover we note that condition~(\ref{0-field_a}) is gauge invariant under any regular gauge transformation
of the unbroken theory. In this sense the zero-field equation is indeed a preferred solution, it is the only
field configuration allowing dynamical symmetry breaking that is gauge invariant as well as duality invariant.
It is missing to describe how the dynamical symmetry breaking is achieved for this particular condition. The only difference
in relation to the previous treatment is that we have to account separately for the regular and non-regular components of the fields.
With out loss of generality let us define the regular field $\tilde{C}^{(0)}$ by subtracting the non-regular part of $A$ from the $\tilde{C}$
field
\be
\tilde{C}^{(0)}_\mu=\tilde{C}-\hat{\epsilon}\,A^{\mathrm{nreg}}\ \Rightarrow\ \tilde{C}^{(0)}_\mu=-\hat{\epsilon}\,A^{\mathrm{reg}}_\mu\ \ .
\lb{0-field_eq}
\ee
The second expression is directly obtained from the non-regular zero-field equation~(\ref{0-field}), hence it constitutes its regular
version, being a particular case of our construction. The respective effective electric action and electromagnetic field definitions are
\be
S^{(0)}_{\mathrm{Electric}}=-\int_M F^{\mathrm{reg}}_{\mu\nu}F^{{\mathrm{reg}}\,\mu\nu}\ \ ,\ \ E^i=2F^{{\mathrm{reg}}\,0i}\ \ ,\ \ B^i=\epsilon^{0ijk}F^{\mathrm{reg}}_{jk}\ ,
\lb{el_eff_0_0}
\ee
given only in terms of the regular fields $A^{\mathrm{reg}}$. We note that the usual normalization is obtained by
considering a rescaling of the fields by a factor of $1/2$. In the presence of sources we can generalize this construction
by including the non-regular field $\varphi^g_\mu$ as introduced in equation~(\ref{varphi}). Then the
generalization of~(\ref{0-field_eq}) is
\be
\tilde{C}^{(0,\varphi)}_\mu=\tilde{C}_\mu-\hat{\epsilon}\,A_\mu^{\mathrm{nreg}}+\hat{\epsilon}\,\varphi_\mu^g\ \Rightarrow\ \tilde{C}^{(0,\varphi)}_\mu=-\hat{\epsilon}\,A^{\mathrm{reg}}_\mu\ .
\lb{0-field_eq_J}
\ee
As already discussed the field $\varphi^g$ is non-regular as expressed by~(\ref{varphi}),
hence cannot be included either in the regular part of $A$ or in the field $a$ which is also regular by construction.
We have shown that duality and gauge invariance of the zero-field equation differentiate it as a
preferred configuration, however this choice is not mandatory. In principle the relation between the $a$ field
and the $A$ field describes as the original degrees of freedom encoded in the $A$ field are mapped to the surviving
degrees of freedom encoded in the $a$ field. If the mechanisms presented in this work exist in real world, it must
be experiments to test this theoretical preference. Also it is important to stress that in this discussion we are
assuming that there is a direct mapping between $a$ and $A$, our generic construction does not require any map
between both fields $a$ and $A$, it simply kills the degrees of freedom encoded in the $C$ field and exchanges
the degrees of freedom encoded in the $A$ field of the $U_e(1)\times U_g(1)$ theory, by the degrees of freedom
encoded in the $a$ field of the effective $U_{\bar{e}}(1)$ theory. Similar results hold for the magnetic case.

\section{\lb{conc}Conclusions}

In this work we have presented a dynamical symmetry breaking mechanism for extended $U_e(1)\times U_g(1)$ electromagnetism
containing both photons $A$ and pseudo-photons $C$. This mechanism renders broken effective $U_{\bar{e}}(1)$ and $U_{\bar{g}}(1)$ theories with
only one gauge field incorporating both electric and magnetic four-currents. The most remarkable feature of our
construction is that the extended singularities (Dirac string~(Dirac) or Wu-Yang fiber bundle~(Wu \& Yang~1975))
characteristic of theories with one single $U(1)$ gauge field containing both electric and magnetic four-currents,
are absent. Both four-currents are still present in the effective theories but are described through corrections to
the standard definitions of the physical electromagnetic fields inherit from the unbroken theory. In the same manner, also the canonical
momenta definitions gain corrections inherited from the unbroken theory. At variational level in the effective broken theories
these corrections to the momenta are justified by properly considering the boundary contributions induced by the dynamical
symmetry breaking mechanism. We have also shown that, in the presence of both electric and magnetic four-currents, trivial field configurations
are not compatible with the equations of motion for both the gauge fields $A$ and $C$. This proves non-triviality of the
theory. Moreover, although extended $U_e(1)\times U_g(1)$ electromagnetism is in complete agreement with the phenomenological Maxwell
equations, these results imply that it is not equivalence to standard variational Maxwell theory.

In this work we have addressed only local current densities which can be coupled, at variational level, directly to the gauge connections.
This construction have previously been considered in~(Schwinger~1951) holding that the action is explicitly gauge invariant.
However it excludes non-local currents such as constant background currents, non-trivial topological charge configurations,
large gauge transformations, Wilson line effects and other topological effects. When such backgrounds are considered,
depending on the specific framework, can be generated a Proca mass for the surviving gauge field~(Schwinger~1962, Anderson~1963, Proca~1988).
This is studied in detail and applied to unmagnetized plasmas in~(Castelo Ferreira \& Mendon\c{c}a~2006, Mendon\c{c}a \& Castelo Ferreira~2006).
We have also shown that there is a preferential field configuration that corresponds to the zero-field equation~(Cabibbo \& Ferrari~1962, Schwinger~1966).
This configuration corresponds to the self-dual point of a duality constituted by a combination of Hodge dualities that
is a global symmetry of the unbroken action. In addition the zero-field condition is also the only gauge field configuration
that, when written explicitly as a non-differential relation between the gauge fields (instead of a condition written in terms
of the connections), is gauge invariant. These characteristics justify the zero-field equation, from a theoretical point
of view, as a preferred field configuration.

In the construction presented in this work the gauge fields regularity in the effective broken theories is accomplished
by a correction to the physical fields and charge definitions inherit from the original unbroken theory instead of being
achieved at effective action level. Although being a correct framework, a more desirable framework, would be to accomplished
this program explicitly at the level of the action by some sort of mechanism. Although not clear to the author how to achieve
this construction, possible approaches may be to consider extra fields. For example scalar fields~(Witten~1979), in which case
the corrections to the electric and magnetic field corrections may be justified as topological charge contributions due to the
cross Hopf-term maintaining simultaneously $P$ and $T$ invariance. Also other possible mechanism could include a boundary
conformal theory that offsets the boundary contributions of gauge transformations, similarly to Abelian NWZW-models~(Wess \& Zumino~1971, Witten~1983).
This assumption is justified by noting that the inherit quantum momenta can be described in terms of boundary contributions as
we have shown. These topics deserve a detailed study somewhere else.

For last we note that our construction is assuming that monopoles are treated in the same footing as electrons do, being
fundamental particles, i.e. Dirac monopoles~(Dirac). Other possible approach to magnetic monopoles is to consider them as gauge
configurations of non-Abelian gauge groups, i.e. t'Hooft-Polyakov monopoles~(T'Hooft~1974, Polyakov~1974). These can still be treated as
particles in effective theories below the non-Abelian gauge symmetry breaking energy and our results
could still be related to this framework. However for the particular case of BPS
dyons~(Julia \& Zee~1975, Prasad \& Sommerfield~1975, Bogomolny~1976) it was shown by
Seiberg and Witten~(Seiberg \& Witten~1994, see also Witten~1995) that a duality between both electric and
magnetic vacuum exists, which is distinct from the duality presented in our work. Furthermore as opposed
to our construction Seiberg-Witten vacuum, besides being $P$ and $T$ violating, is chiral. Although not
completely clear to the author at this stage, this fact may render both approaches non-equivalent, or at
least mean that both approaches are describing different physical systems (at least concerning chiral
symmetry breaking which is usually associated with the inclusion of fermionic effects).
Concerning our approach and pseudo-photons, it is important to stress that the physical
degrees of freedom are encoded in a gauge vector field (photon) and gauge pseudo-vector field (pseudo-photon),
as opposed to other approaches where two gauge vector fields are considered to correspond to one physical
photon and one auxiliary non-physical photon which is integrated out of the theory. Also for pseudo-photon theories, exist
already available theoretical results, which indicate that may exist in low-energy regimes, both physical systems which
exhibit an unbroken phase~(Castelo Ferreira~2006), physical systems which exhibit a broken electric
phase~(Castelo Ferreira \& Mendon\c{c}a~2006, Mendon\c{c}a \& Castelo Ferreira~2006) and systems that
exhibit a broken magnetic phase~(Castelo Ferreira~2007). Also, these works indicate that, the pseudo-photon may be a truly
physical field, instead of a mathematical auxiliary field.

{\bf Acknowledgements}
This work was supported by SFRH/BPD/17683/2004.


\begin{thebibliography}{90}

\bibitem[Anderson~1963]{Anderson} P. W. Anderson, Phys. Rev.  {\bf 130} (1963) 439-442.
\bibitem[Berkovits~1996]{Berkovits} N. Berkovits, Phys. Lett.  {\bf B395} (1997) 28-35, \texttt{hep-th/9610134}.
\bibitem[Bogomolny~1976]{Bogomolny} E.B. Bogomolny, Sov. J. Nucl. Phys. {\bf 24} (1976) 449; Yad. Fiz. {\bf 24} (1976) 861.
\bibitem[Brandt \& al.~1979]{Brandt} R. A. Brandt, F. Neri and D. Zwanzinger, Phys. Rev. {\bf D19} (1979) 1153-1167.
\bibitem[Cabibbo \& Ferrari~1962]{Cabibbo} N. Cabibbo and E. Ferrari, Il Nuovo Cimento {\bf XXIII} (1962) 1147-1154.
\bibitem[Cardoso de Mello \& al.~1996]{Cardoso} P. C. R. Cardoso de Mello, S. Carneiro e M. C. Nemes, Phys. Lett. {\bf B384} (1996) 197-200, \texttt{hep-th/9609218}.
\bibitem[Carneiro~1997]{Carneiro} S. Carneiro, JHEP 9807 (1998) 022, \texttt{hep-th/9702036}.
\bibitem[Castelo Ferreira~2005]{PCF_2005a} P. Castelo Ferreira, J. Math. Phys. {\bf 47} (2006) 072902, \texttt{hep-th/0510063}.
\bibitem[Castelo Ferreira \& Mendon\c{c}a~2006]{PCF_2005b} P. Castelo Ferreira and J. T. Mendon\c{c}a, \texttt{hep-th/0601171}.
\bibitem[Castelo Ferreira~2006]{PCF_2006} P. Castelo Ferreira, accepted for publication in Phys. Rev. {\bf B}, hep-ph/0609239.
\bibitem[Castelo Ferreira~2007]{PCF_2007} P. Castelo Ferreira, \texttt{hep-th/0703193}; \texttt{hep-th/0703194}.
\bibitem[Dirac]{Dirac} P. A. M. Dirac, Proc. Roy. Soc. {\bf A133} (1931) 60; Phys. Rev. {\bf 74} (1948) 817.
\bibitem[Girvin \& MacDonald~1987]{Girvin}   S. M. Girvin and A. H. MacDonald, Phys. Rev. Lett. {\bf 58} (1987) 1252.
\bibitem[Jackson~1975]{Jackson} J. D. Jackson, \textit{Classical Electrodynamics}, $2^{nd}$ Edition, John Wiley \& Sons, 1975.
\bibitem[Jain~1989]{Jain}     J. K. Jain, Phys. Rev. Lett. {\bf 63} (1989) 199.
\bibitem[Julia \& Zee~1975]{Julia} B. Julia and A. Zee, Phys. Rev. {\bf D11} (1975) 2227.
\bibitem[Laughlin~1982]{Laughlin}  R. B. Laughlin, Phys. Rev. Lett. {\bf 50} (1982) 1395.
\bibitem[Mendon\c{c}a \& Castelo Ferreira~2006]{Mendonca} J. T. Mendon\c{c}a and P. Castelo Ferreira, Europhys. Lett. {\bf 75} (2006) 189, \texttt{hep-th/0601166}.
\bibitem[Polyakov~1974]{Polyakov} A. M. Polyakov, JETP Lett. {\bf 20} (1974) 194; Pisma Zh. Eksp. Teor. Fiz. {\bf 20} (1974) 430.
\bibitem[Prasad \& Sommerfield~1975]{Prasad} M. K. Prasad and C. M. Sommerfield, Phys. Rev. Lett. { \bf 35} (1975) 760.
\bibitem[Proca~1988]{Proca} \textit{Alexandre Proca (1897-1955): Scientific Publications}, G. A. Proca, 1988.
\bibitem[Schwinger~1951]{Schwinger_1951} J. Schwinger, Phys. Rev. {\bf 82} (1951) 664-679.
\bibitem[Schwinger~1962]{Schwinger_1962} J. Schwinger, Phys. Rev.  {\bf 125} (1962) 397-398.
\bibitem[Schwinger~1966]{Schwinger_1966} J. Schwinger, Phys. Rev. {\bf 144} (1966) 1087.
\bibitem[Seiberg \& Witten~1994]{SW} N. Seiberg and E. Witten, Nucl. Phys. {\bf B426} (1994) 19, Erratum-ibid. {\bf B430} (1994) 485-486, \texttt{hep-th/9407087}; Nucl. Phys. {\bf B431} (1994) 484, \texttt{hep-th/9408099}.
\bibitem[Singleton~1995]{Singleton} D. Singleton, Int. J. Theor. Phys. {\bf 35} (1996) 2419-2426, \texttt{hep-th/9509157}; Int. J. Theor. Phys. {\bf 34} (1995) 2453, 
\bibitem[T'Hooft~1974]{THooft} G. 't Hooft, Nucl. Phys. {\bf B426} (1974) 276.
\bibitem[Tsui \& al.~1982]{Tsui} D. C. Tsui, H. L. Stormer and A. C. Gossard, Phys. Rev. Lett. {\bf 48} (1982) 1559.
\bibitem[Wess \& Zumino~1971]{WZ} J. Wess, B. Zumino, Phys. Lett. {\bf B37} (1971) 95.
\bibitem[Witten~1979]{Witten_1979} E. Witten, Phys. Lett. {\bf B86} (1979) 283-287.
\bibitem[Witten~1983]{Witten_1983} E. Witten, Nucl. Phys. {\bf B223} (1983) 422-432.
\bibitem[Witten~1995]{Witten_1995} E. Witten, Selecta Math. {\bf 1} (1995) 383, \texttt{hep-th/9505186}.
\bibitem[Wu \& Yang~1975]{WY} T. T. Wu and C. N. Yang, Phys. Rev. {\bf D12} (1975) 3845; Phys. Rev. {\bf D14} (1976) 437
\texttt{hep-th/9701040}.
\bibitem[Zhang \& al.~1989]{Zhang}    S.C. Zhang, T. H. Hansson and S. Kivelson, Phys. Rev. Lett. {\bf 62} (1989) 82.
\bibitem[Zwanzinger~1968]{Zwanzinger} D. Zwanzinger, Phys. Rev. {\bf 176} (1968) 1489-1495; Phys. Rev. {\bf D3} (1971) 880-891.


\end{thebibliography}
\end{document}